\documentclass[12pt]{article}
\usepackage{amsmath}
\usepackage{graphicx}
\usepackage{geometry}
\usepackage[applemac]{inputenc}

\ifx\pdfoutput\@undefined\usepackage[usenames,dvips]{color}
\else\usepackage[usenames,dvipsnames]{color}
\IfFileExists{pdfcolmk.sty}{\usepackage{pdfcolmk}}{} 
\fi
\usepackage{hyperref}
\usepackage[all]{hypcap}

\begin{document}

\title{Harnessing the Complexity of Education\\ with Information Technology} 

\author
{Carlos Gershenson$^{1,2}$\\
$^{1}$ Departamento de Ciencias de la Computaci\'on\\
Instituto de Investigaciones en Matem\'aticas Aplicadas y en Sistemas \\
Universidad Nacional Aut\'onoma de M\'exico\\
A.P. 20-126, 01000 M\'exico D.F. M\'exico\\
Tel. +52 55 56 22 36 19 \
Fax +52 55 56 22 36 20 \\
\href{mailto:cgg@unam.mx}{cgg@unam.mx} \
\url{http://turing.iimas.unam.mx/~cgg} \\
$^{2}$ Centro de Ciencias de la Complejidad \\
Universidad Nacional Aut\'onoma de M\'exico
}


\maketitle

Education at all levels is facing several challenges in most countries~\cite{Linn1987SciEdu,Robinson1998All-Our-Futures,Hines2013EduChallenges,Kremer2013EduDevWorld}, such as low quality, high costs, lack of educators, and unsatisfied student demand. Traditional approaches are becoming unable to deliver the required education. Several causes for this inefficiency can be identified. I argue that beyond specific causes, the lack of effective education is related to \emph{complexity}. However, information technology is helping us overcome this complexity.

Complexity can be measured with information theory and can be seen as the balance between stability and variability~\cite{Langton1990,Prokopenko:2008,GershensonFernandez:2012}: phenomena without change or with constant change cannot exhibit complex behavior. It has been noted that in order to actively control a complex system, the controller has to be at least as complex as the controlled~\cite{Ashby1956,BarYam2004}
. For example, a successful healthcare provider has to match the complexity of the patients she attends. Treatment is highly specific for different patients, so a general practitioner must have a high complexity to attend patients with diverse conditions. Concerning most preventive services, these are similar for most patients and thus can be delivered efficiently by providers with a lower complexity~\cite{BarYamAJPH2006}. A similar approach can be used to study education and its complexity: a successful educational system has to match the complexity of its students.

Traditional, teacher-based education is limited by the complexity of the teacher. Even the best teachers have finite abilities, in the sense that they can only deliver content in a limited number of ways to a diverse group of students. As the group grows in size and diversity, its complexity will increase, demanding a greater teacher complexity or limiting the complexity of the class. Different solutions which have been applied to overcome this inherent limitation are illustrated in Fig. 1 and can be classified by:

\begin{description}
\item [Quantity.] Reduce the student/teacher ratio by having smaller classes or more teachers. This increases the cost of education.
\item [Quality.] Reduce the diversity of students in each class, which become specialized. Thus, students with similar abilities are able to absorb similar content at similar speeds. This requires a large institutional effort and is restricted to densely populated areas.
\item [Scope.] Reduce the complexity of content, i.e. standardize. This reduces the knowledge which diverse students can absorb. Moreover, faster learners become bored and slower learners become frustrated~\cite{Csikszentmihalyi1990Flow}.
\end{description}

\begin{figure}[htbp]
\begin{center}
  \includegraphics[width=0.9\textwidth]{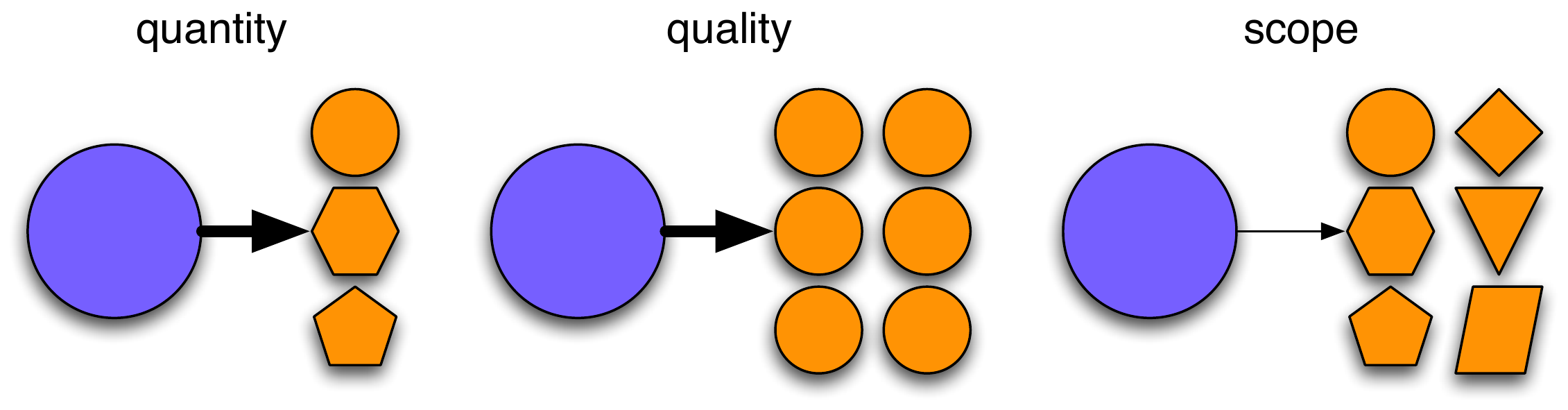}\\
\caption{Traditional solutions to the complexity limits of educators (purple) depending on the combined complexity of their students (orange): decrease the student/teacher proportion with smaller classes, reduce diversity of groups, and decrease the scope of the content delivered.}
\label{fig:cx}
\end{center}
\end{figure}

None of these approaches is satisfactory for the challenges education is facing. Increasing motivation or budgets is positive, but only palliative, as none of these increase the required complexity of the educational system, which is the main limitation for improving education.
For example, India estimates that it will require one thousand new universities by 2020~\cite{Davidson20101000-new-univer}. This is unfeasible with a traditional educational system.

Nevertheless, information technology is allowing us to go beyond the complexity of the teacher and beyond the complexity of the school. 
 Traditional approaches have been reducing the complexity of the students. Technology is allowing us to increase the complexity of the teaching system, potentially harnessing the complexity of hundreds of institutions and thousands of teachers to deliver education to millions of students.

The strict difference between teacher delivering education and student receiving education is changing towards a more distributed framework, where teachers and students form a co-learning community. Technology is serving as a mediator, coordinating efforts and demands of agents in an increasingly global educational system. This shift can be compared with the phenomenon of Wikipedia~\cite{Wikipedia}. Instead of having a dedicated company coordinating high complexity authors to publish an encyclopedia, information technology has allowed the development of a distributed system where the aggregated complexity of thousands of authors of diverse complexities has overcome the most comprehensive private encyclopedias.

For example, the ``Skype Grannies" project allowed retired people mainly from the UK to mediate the education of children in rural India~\cite{Clark2011SkypeGrannies}. In specific cases, technology can facilitate learning even without adult guidance, as older children become facilitators of younger ones exploiting available content~\cite{mitra2005holeInTheWall}.

The increasing availability of free online content is allowing more independent learning paths. Wikipedia, online libraries, and sites such as Khan Academy~\cite{KhanAcademy} provide material for students and teachers which effectively can increase the complexity of the education delivered.

As for higher education, massive online open courses (MOOCs) have become very popular in recent months~\cite{pappano2012year}, effectively delivering educational content to millions of students from all over the world. This content is not personalized, but students select their courses according to their interests, so there is a natural match between the offered material and the student requirements.

These novel teaching tools are not to replace the current educational system, based on teachers and schools. These tools are transforming the educational system. They empower teachers with a greater complexity to deliver a better education.

Different teaching resources coordinated by teachers can have a greater complexity than millions of students, thus opening the possibility of delivering education in a much more effective and efficient way. There is an analogy with insect colonies: individual insects have a relatively low complexity. However, coordinating through their environment~\cite{TheraulazBonabeau1999}, insect colonies are able to complete tasks which are much more complex than any individual. Technology is having a similar role for coordinating our individual efforts~\cite{Bernstein2012GlobalBrain} and increasing the complexity of the educational system. Technology not only allows to have instant communication and personalized content delivered individually, but it is turing the complexity limitation of education on its back, as illustrated in Fig. 2:

\begin{description}
\item [Quantity.] Thousands of teachers can interact with millions of students. Students become also teachers in learning communities. Each individual in the system can learn and potentially generate novel content. 
\item [Quality.] Diverse students can obtain different content, suited to their abilities and requirements.
\item [Scope.] Content is personalized. With a broad diversity of content to choose from, this can be selected to maximize student learning and satisfaction.
\end{description}

\begin{figure}[htbp]
\begin{center}
  \includegraphics[width=0.7\textwidth]{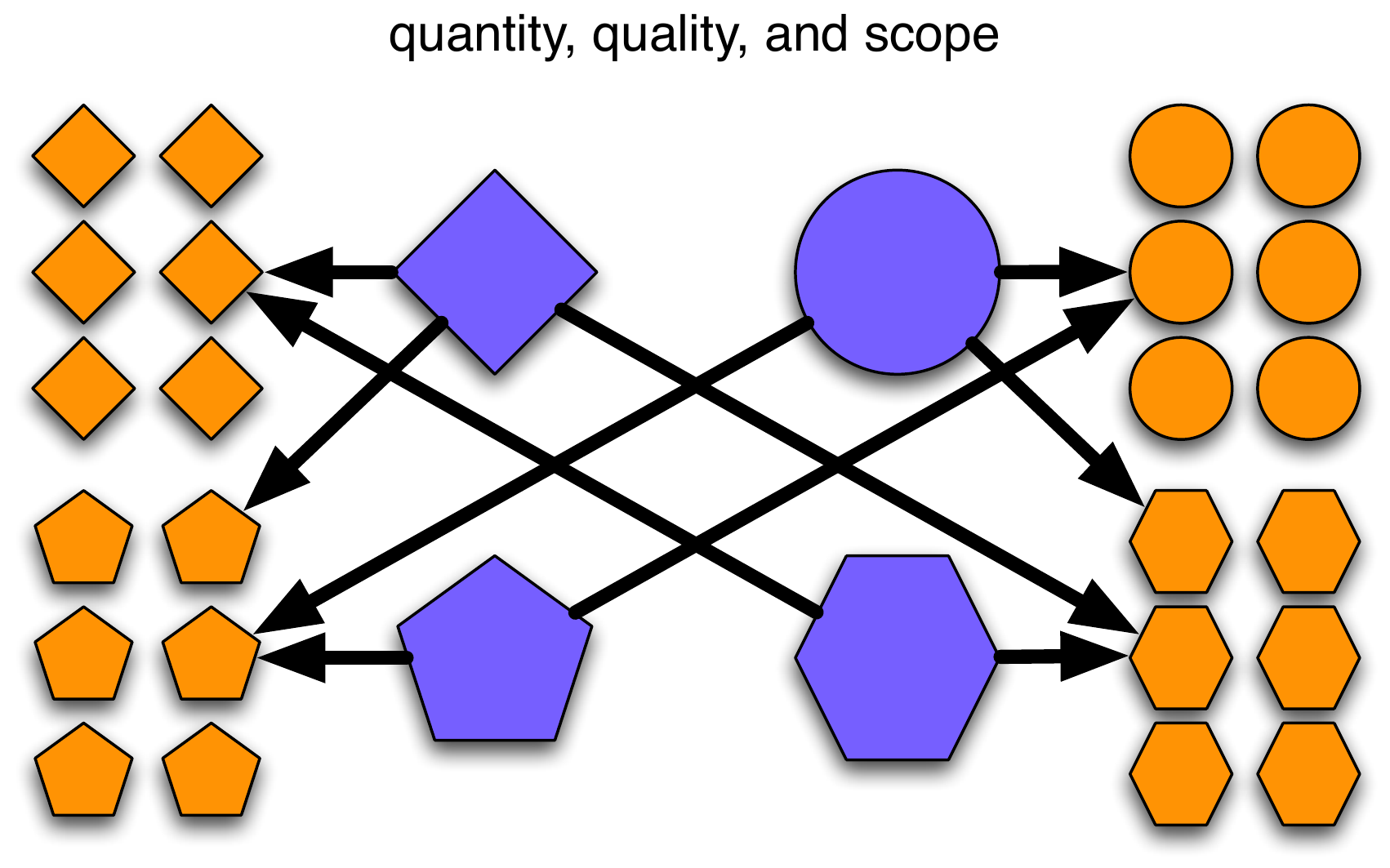}\\
\caption{Technology allows a complexity increase of the educational system, rising the quality, quantity, and scope of the education delivered.}
\label{fig:cx2}
\end{center}
\end{figure}

Even when information technology offers promising improvements for our educational systems, several economic, political, social, and cultural challenges lie ahead. One of them is language, although communities are providing translation of educational content and English is becoming a \emph{lingua franca}.
Still, the complexity increase that technology is allowing offers many opportunities. For example, education is correlated with several well-being indicators, such as life expectancy, less children per woman, higher incomes and less crime~\cite{Gapminder}. Correlation is not causation, but few would disagree that improving the education of the world is a necessary step in attending global problems. 
The benefits of a global distributed educational system will also be global. 

The persistence of knowledge in writing marked the transition between prehistory and history. The mass production of texts with the printed press was one of the main factors leading to the scientific revolution. As technology increases our capacity to store, transmit, and access knowledge, how will this transform our globalized culture?

\section*{Acknowledgments}

I should like to thank Daphne Koller for useful comments. 
This work was partially supported by SNI membership 47907 of CONACyT, Mexico. 

\bibliographystyle{unsrt}
\bibliography{carlos,edu,complex,RBN,sos,COG}

\end{document}